# Multi-Layer Perceptron-Based Relay Node Selection for Next-Generation Intelligent Delay-Tolerant Networks

**Zhekun Huang**

School of Computer Science

University of Nottingham, England

psxzh7＠nottingham．ac．uk

**Milena Radenkovic**

School of Computer Science

University of Nottingham, England

milena．radenkovic＠nottingham．ac．uk

Contents

# Abstract

Delay Tolerant Networks (DTNs) are critical for emergency communication in highly dynamic and challenging scenarios characterized by intermittent connectivity, frequent disruptions, and unpredictable node mobility. While some protocols are widely adopted for simplicity and low overhead, their static replication strategy lacks the ability to adaptively distinguish high-quality relay nodes, often leading to inefficient and suboptimal message dissemination.

To address this challenge, we propose a novel intelligent routing enhancement that integrates machine learning-based node evaluation into the Spray and Wait framework. Several dynamic, core features are extracted from simulation logs and are used to train multiple classifiers - Multi-Layer Perceptron (MLP), Support Vector Machine (SVM), and Random Forest (RF) - to predict whether a node is suitable as a relay under dynamic conditions.

The trained models are deployed via a lightweight Flask-based RESTful API, enabling real-time, adaptive predictions. We implement the enhanced router MLPBasedSprayRouter, which selectively forwards messages based on the predicted relay quality. A caching mechanism is incorporated to reduce computational overhead and ensure stable, low-latency inference.

Extensive experiments under realistic emergency mobility scenarios demonstrate that the proposed framework significantly improves delivery ratio while reducing average latency compared to the baseline protocols. Among all evaluated classifiers, MLP achieved the most robust performance, consistently outperforming both SVM and RF in terms of accuracy, adaptability, and inference speed. These results confirm the novelty and practicality of integrating machine learning into DTN routing, paving the way for resilient and intelligent communication systems in smart cities, disaster recovery, and other dynamic environments.

Chapter 1

# 1. Introduction

Delay-Tolerant Networks (DTNs) have emerged as a novel and resilient communication paradigm to tackle the challenging nature of intermittent connectivity, high latency, and frequent disconnections in dynamic and opportunistic environments [1]. By employing the store-carry-forward mechanism, DTNs enable message delivery across sparse, delay-prone topologies, rendering them highly suitable for vehicular networks, rural deployments, and time-critical emergency response systems.

Among DTN routing protocols, Spray and Wait [2] remains a seminal yet inherently static method. Its fixed replication strategy offers a balance between delivery efficiency and resource utilization but fails to dynamically adapt to context-aware challenges in heterogeneous and evolving environments. This limitation has motivated the development of adaptive and intelligent routing strategies that exploit contextual features, mobility dynamics, and data-driven decision-making.

The integration of machine learning (ML) into networking has provided novel opportunities to address these challenges, revolutionizing resource allocation, traffic prediction, and anomaly detection across diverse communication systems. In mobile ad hoc networks (MANETs) and Internet-of-Things (IoT) deployments, ML-driven models have been applied to enhance routing, detect faults, and mitigate congestion [3]. These approaches illustrate the potential of ML to autonomously adapt to dynamic, non-stationary, and heterogeneous network conditions. Particularly in infrastructure-less and delay-prone scenarios such as disaster recovery zones, conflict areas, or remote agricultural regions, ML provides context-aware decision-making capabilities that are nearly impossible to hardcode.

The critical importance of DTNs has been further underscored by global crises. For example, during the COVID-19 pandemic, conventional communication infrastructureswere frequently overloaded, giving rise to novel UAV-based and mobile DTNs for diagnostics and remote message relaying. Similarly, in natural disasters and conflict zones, DTNs have supported first responders, drones, and mobile healthcare units under highly dynamic and uncertain communication opportunities. These real-world challenges underscore the urgent need for intelligent and adaptive routing mechanisms.

The feasibility of ML in DTNs has been reinforced by advances in edge intelligence, model compression, and low-latency inference techniques. Lightweight neural networks such as shallow MLPs, decision tree ensembles, and federated learning frameworks now enable on-device, context-aware inference under stringent constraints of energy and bandwidth [4]. This evolution creates novel opportunities for dynamic and scalable routing strategies that align with DTNs' decentralized nature.

Recent studies highlight the promising potential of ML to overcome routing challenges. For instance, Radenkovic et al. [5] enhanced emergency communications with a Random Forest classifier in smart cities, while Wang et al. [6] demonstrated an energy-aware, ML-driven strategy capable of balancing power consumption with reliability. These contributions illustrate how ML introduces novel and practical solutions to real-world DTN challenges.

Moreover, supervised learning approaches for dynamic node classification have shown effectiveness. Mtibaa et al. [7] introduced people-centric, socially-aware routing, while Taleb and Ksentini [8] proposed mobility-driven service migration (Follow-Me Cloud). By extracting temporal-spatial mobility patterns and contact dynamics, these works highlight the challenge and necessity of predicting high-quality relay nodes in dynamic DTN environments.

Building upon these insights, we propose a novel hybrid framework that integrates external ML classifiers—Multilayer Perceptron (MLP), Support Vector Machine (SVM), and Random Forest (RF)—into an enhanced Spray and Wait protocol within the ONE Simulator. Our design dynamically extracts features such as contact frequency, node degree, contact duration, hop count, and delivery delay, and transmits them via lightweight HTTP requests to a Flask-based inference API. The API performs real-time binary classification to determine whether a node should act as a relay.

Our evaluation demonstrates that the proposed MLP-based router (MLPBased-SprayRouter) consistently outperforms both baseline Spray and Wait and alternative ML variants in terms of delivery ratio, latency, and routing overhead. This work highlights the capability of combining lightweight feature extraction with runtime ML prediction, thereby establishing a dynamic, context-aware, and resilient routing paradigm for DTNs in urban, rural, and emergency scenarios.

Chapter 2

# 2. Literature Review

Delay-Tolerant Networks (DTNs) have been proposed to support communication in intermittently connected or infrastructure-less environments. Traditional routing protocols such as Epidemic Routing [9] and Spray and Wait [2] laid the foundational strategies by either flooding messages or controlling message copies to improve delivery rates and reduce overhead. However, these approaches often overlook energy efficiency and contextual adaptability, making them suboptimal for large-scale or resource-constrained deployments.

Pelusi et al. [10] emphasized the importance of context-aware and opportunistic forwarding strategies in DTNs. Building upon this, social-aware and history-based methods have been proposed to incorporate contact frequency, mobility similarity, and node centrality into relay selection. For instance, the Prophet protocol improves upon Epidemic Routing by estimating delivery predictability based on past encounters, yet remains heuristic in nature.

To overcome the limitations of handcrafted strategies, machine learning (ML) has emerged as a promising tool in DTN routing. Ashapu et al. [11] provided a comprehensive review of ML applications in DTNs, outlining supervised, unsupervised, and reinforcement learning strategies for adaptive routing. The authors identified the potential of ML to learn non-linear decision boundaries, adapt to dynamic topologies, and generalize across mobility patterns.

Recent studies have particularly focused on supervised ML models for classifying relay nodes. For instance, Min Wook Kan et al. [12] proposed a hybrid energy-aware scheme integrating ML prediction into routing logic. Sharma et al. [13] developed MLProph, which uses historical contact and delivery statistics to predict relay quality. These models often rely on features such as contact frequency, inter-contact time, and average delivery latency, reflecting a growing consensus that data-driven classification can outperform static threshold-based rules.

Radenkovic et al. [5] demonstrated that Random Forest classifiers embedded within routing protocols significantly enhanced delivery success rates in smart city emergencies. Their approach involved training ensemble models on node mobility, density, and delivery statistics, showing measurable gains in delivery probability and latency reduction. Similarly, Wang et al. [6] proposed a clustering-based approach with energy-delay trade-offs, revealing the feasibility of joint learning and optimization in DTN routing.

Beyond these, several state-of-the-art works have explored AI-driven congestion control, caching, and heterogeneous network management. Radenkovic and Grundy [14] introduced congestion-aware dissemination in social opportunistic networks. Radenkovic and Huynh [15] proposed multi-agent deep reinforcement learning for cognitive caching at the network edge, while Radenkovic et al. [16] enabled real-time communication in heterogeneous drone–vehicle networks. Huynh and Radenkovic [17] further advanced multi-layer spatio-temporal caching in mobile edge and fog systems. These studies highlight the multi-criteria and multi-modal complexity of DTNs, addressing congestion, caching, and service orchestration. However, their focus differs from our work, which tackles the more fine-grained problem of relay selection inside the Spray and Wait protocol.

Beyond decision trees and SVMs, neural network-based models such as Multi-Layer Perceptrons (MLP) are gaining traction due to their ability to model non-linear interactions between features. While more computationally intensive than simpler models, MLPs can be effectively deployed with lightweight inference engines, especially when coupled with optimized runtime environments like Flask APIs. Biau and Scornet [18] further reinforced the effectiveness of ensemble methods in structured data tasks, such as node behavior modeling in DTNs.

Moreover, the relevance of deep learning techniques is also emerging. Kattenborn et al. [19] reviewed convolutional neural networks (CNNs) in spatial inference tasks, while Li et al. [20] explored LSTM architectures for time-dependent sequence modeling. Though these approaches are promising, their computational cost currently limits their adoption in resource-constrained DTN environments.

In terms of experimental support, the ONE Simulator remains the de facto standard for testing DTN protocols [21]. Comparative studies using ONE have benchmarked routing schemes under varying mobility models and message generation patterns [22], and it remains highly extensible for integration with external ML services.

Our work builds upon these developments by comparing three supervised classifiers—MLP, SVM, and Random Forest—trained on mobility and delivery features derived from ONE Simulator logs. By deploying the trained MLP model via a Flask API and integrating it into the modified Spray and Wait protocol (MLPBasedSprayRouter), we enable real-time relay quality prediction. The proposed system improves delivery probability and reduces latency in both weekday and holiday urban scenarios, thereby demonstrating the practical viability of ML-enhanced DTN routing.

Chapter 3

# 3. Methodology

## 3.1. System Design and Architecture

### 3.1.1. System Architecture

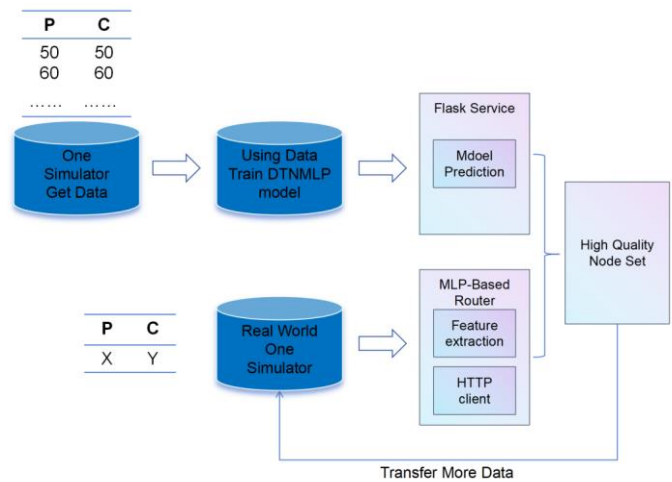

Figure 3.1: Overall system architecture integrating ONE Simulator, MLP model, and routing logic

The proposed system architecture, illustrated in Figure 3.1, is designed to integrate machine learning-based decision-making into Delay Tolerant Network (DTN) routing via simulation in the ONE Simulator environment. The architecture is divided into two major phases: an offline model training phase and an online routing decision phase.

In the offline phase, simulation data is generated using the ONE Simulator under a variety of controlled scenarios defined by parameters such as the number of pedistrains (P) and cars (C). From this data, we extract a set of node-level features that have been empirically found to influence routing effectiveness: contact frequency, node degree, average contact duration, average hop count, average delivery time, number of times a node served as a relay, and number of times it served as a destination [6, 5]. These features are then used to construct a labeled dataset identifying whether a node qualifies as a high-quality relay. Using this dataset, we train a Multilayer Perceptron (MLP) model, which has shown superior classification accuracy compared to SVM and Random Forest alternatives in our evaluation.

In the online phase, the trained MLP model is deployed as a Flask-based HTTP service. A customized routing protocol, named MLPBasedSprayRouter, is implemented within the ONE Simulator to support real-time feature extraction and decision-making. When a message is generated and a forwarding opportunity arises, the router module computes the relevant features for candidate nodes and queries the Flask service through an asynchronous HTTP POST request. The service returns a binary classification indicating whether the candidate is a high-quality node. If so, the message is forwarded accordingly. This approach ensures that only promising relay candidates are selected, reducing unnecessary transmissions and improving delivery efficiency.

This architecture builds on prior work in hybrid intelligent routing , while introducing a modular and extensible framework for integrating external predictive models into simulation-based DTN environments. By decoupling model training and prediction from simulation execution, we enable future extensions such as online learning, ensemble models, or adaptive retraining based on environmental drift.

### 3.1.2. System Execution Flow

The overall execution flow of our DTN routing system involves a closed-loop interaction between data generation, machine learning model training, and real-time routing decision making. As illustrated in Figure 3.2, the process begins with the **One Simulator** generating mobility traces and contact data under a specific scenario configuration (e.g., 50 pedestrians, 60 cars). This data includes detailed information such as contact frequency, node degree, average contact duration, and message delivery statistics. The next step involves **feature extraction and labeling**, where each node's statistical features are computed from the simulation log and labeled as high- or low-quality based on their delivery efficiency, hop count, or frequency of being selected as a relay. These features serve as the training dataset for the classification model.

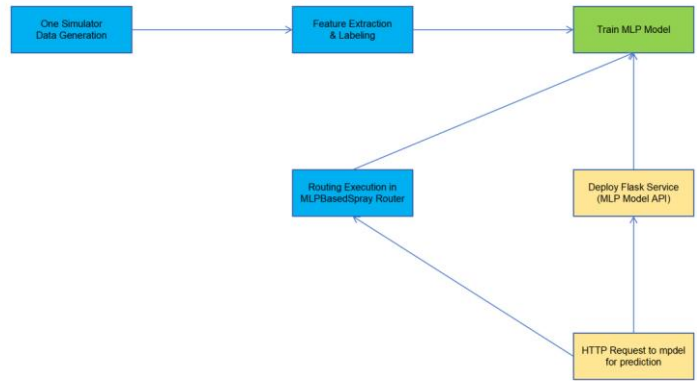

Figure 3.2: System Execution Flow integrating One Simulator and Flask-based MLP classifier

Subsequently, a **Multi-Layer Perceptron (MLP)** classifier is trained using the extracted features. Once the model demonstrates satisfactory classification accuracy, it is deployed via a lightweight **Flask API server**. This allows the trained MLP model to be queried by external routing agents for predictions during live simulation.

In the simulation runtime, our customized **MLPBasedSprayRouter** replaces the default SprayAndWaitRouter in ONE Simulator. During message forwarding, each node performs feature extraction in real time and submits an **HTTP request** to the deployed Flask API. The API returns a binary prediction indicating whether the candidate node qualifies as a high-quality relay. Only those predicted as high-quality are selected for message replication.

This loop continues, and over time, the MLP model can be retrained with updated data to reflect evolving network conditions and node behaviors.

This architecture provides a hybrid design that combines offline learning with online decision making, leading to more adaptive and intelligent DTN routing compared to traditional heuristics. The modularity of the system allows it to be extended with other classifiers (e.g., SVM, Random Forest) or integrated into alternative simulation platforms.

## 3.2. Feature Extraction and Dataset Construction

### 3.2.1. Selected Features and Labeling Strategy

We build upon prior DTN–ML research that has explored the use of features such as contact history, buffer status, node mobility, and delivery latency to learn dynamic relay behavior in opportunistic and space networks [23, 24]. In particular, studies have shown that metrics such as contact duration, buffer occupancy, and link throughput strongly correlate with forwarding effectiveness under highly dynamic and resource-constrained conditions [24].

Motivated by these insights and the challenge of capturing context-aware relay quality, we designed a novel set of representative features dynamically extracted from simulation logs generated by the ONE Simulator. Specifically, two internal reports—ConnectivityDtnsim2R and DeliveredMessagesReport—were utilized to derive features that capture temporal, topological, and delivery-related dynamics.

This feature design strategy aligns with and extends recent context-aware relay selection models such as CARL-DTN [25], which leverage social-tie strength and deliv- ery performance within predictive frameworks. By incorporating both contact-based and delivery-centric features, our approach provides a more adaptive and comprehensive foun- dation for intelligent relay selection in dynamic DTN environments.

We derived the following node-level features:

- **Contact Frequency** ($f$): Total number of contact events involving the node.

- **Degree** ($d$): Number of unique nodes encountered.

- **Average Contact Duration** ($t_{contact}$): Mean duration across all contacts.

- **Average Hop Count** ($h_{avg}$): Mean number of hops for messages relayed.

- **Average Delivery Time** ($t_{delay}$): Mean delivery delay for relayed messages.

- **Relay Count** ($r$): Number of times the node acted as an intermediate relay.

- **Destination Count** ($d_r$): Number of times the node was the message destination.

Inspired by SPR-based classifiers [26], we adopt a history-driven scoring approach. Each node is assigned a score by aggregating its normalized feature values (optionally weighted). A binary label is then assigned based on a median split: nodes above the median are labeled as 1 (high-quality), and those below as 0. This forms the basis for a binary classification task.

```
@0.10 p8 <-> p18 up
@0.10 p39 <-> c71 up
@0.10 p5 <-> p39 up
@0.10 p5 <-> c71 up
@0.10 p46 <-> c91 up
@0.10 c56 <-> c59 up
@0.10 c65 <-> c85 up
@0.10 c61 <-> c84 up
@0.50 c56 <-> c59 down
```

Figure 3.3: Excerpt from ConnectivityDtnsim2Report: Each line represents a contact event between two nodes (e.g., c54 <-> c88). These records form the basis for computing contact frequency and duration.

```
# time ID size hopcount deliveryTime fromHost toHost remainingTtl isResponse path
2052.1000 AC9 642672 5 1780.1000 a110 p13 270 N a110->c89->c102->c105->c63->p13
2225.4000 AC29 550456 3 1361.4000 a110 p10 277 N a110->c89->c102->p10
2396.3000 AC32 730848 5 1449.3000 a110 c77 275 N a110->c102->c105->c72->c63->c77
2564.1000 AC26 706959 1 1785.1000 a110 c71 270 N a110->c71
2566.7000 AC21 932664 9 1944.7000 a110 c87 267 N a110->c102->c106->p8->c72->p24->c64->p9->c62->c87
3086.3000 AC97 745697 1 229.3000 a110 c66 296 N a110->c66
3226.2000 AC39 877158 5 2074.2000 a110 c84 265 N a110->c71->c65->p39->c74->c84
```

Figure 3.4: Excerpt from DeliveredMessagesReport: Each entry logs a successful delivery, along with hop count, delay, and routing path.

### 3.2.2. Feature Distribution Analysis

To better understand the discriminative power of selected features, we visualize their statis- tical distributions for both high-quality (Label=1) and low-quality (Label=0) relay nodes. Figure 3.5 presents a series of kernel density plots for all major raw and normalized features in the dataset.

Key observations include:

- **Contact Frequency and Node Degree:** High-quality nodes exhibit significantly higher values in both contact_freq and degree, indicating stronger centrality and social connectivity.

- **Hop Count and Delivery Time:** Features such as avg_hop_count and avg_delivery_time are lower among high-quality nodes, confirming their higher forwarding efficiency.

- **Relay and Destination Participation:** The distributions of as_relay_count and as_destination_count suggest that high-quality nodes are more likely to serve as intermediaries in message forwarding.

- **Score Feature:** The score attribute shows a slightly longer tail for high-quality nodes, indicating that a small subset consistently achieves higher routing impact.

These insights support the feasibility of using these features in a supervised learn- ing framework and justify the use of classifiers such as MLP, SVM, and Random Forest, which can capture nonlinear patterns and complex decision boundaries.

### 3.2.3 Dataset Statistics

The full dataset was constructed by aggregating simulation outputs from nine mobility configurations (e.g., 50_60, 60_80), representing varied ratios of pedestrian and vehicular nodes. Each scenario produces a partial dataset, which is concatenated to form a unified training set.

The processed dataset contains the following fields for each node instance:

- contact_freq, degree, avg_contact_duration
- avg_hop_count, avg_delivery_time
- as_relay_count, as_destination_count, label

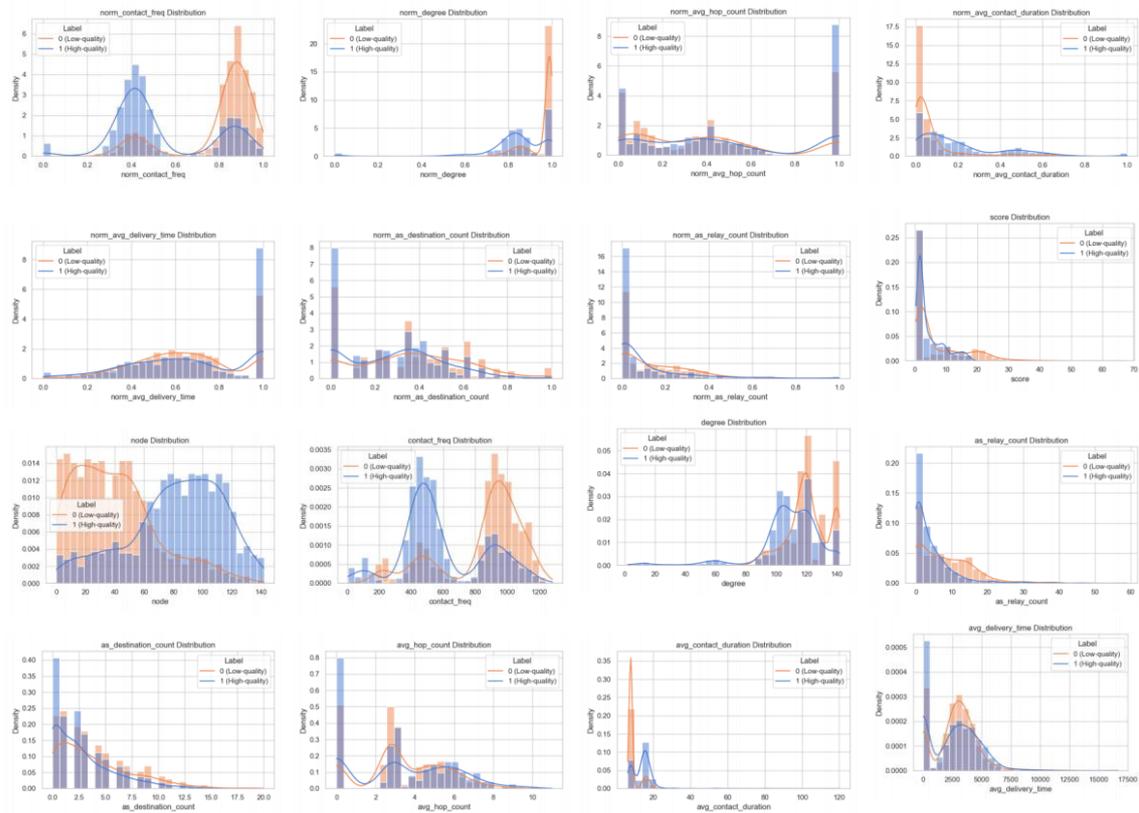

Figure 3.5: Distribution of features across

### 3.2.4. Data Preprocessing

To ensure consistent feature scaling, all numerical features were normalized to the [0, 1] range using Min-Max normalization. For features negatively correlated with relay quality—such as avg_hop_count and avg_delivery_time—reverse normalization was applied to maintain interpretability.

Before model training, node identifiers were removed or encoded as integers. Feature standardization (z-score normalization) was applied where required by certain classifiers. Since labels were generated using a median split, the dataset exhibits balanced class proportions, and no further resampling or weighting techniques were applied.

These preprocessing steps enable robust and efficient model training without introducing class bias or scale-induced instability.

## 3.3. Machine Learning Models and Hyperparameter Tuning

We frame relay selection as a binary classification task: high-quality relay (1) vs. low-quality relay (0). Based on prior DTN-ML research, we evaluate three representative supervised classifiers that balance accuracy, interpretability, and computational feasibility:

- **Multi-Layer Perceptron (MLP)**: A feedforward neural network with two hidden layers (128 and 64 neurons, ReLU activation) trained with Adam and cross-entropy loss. MLPs capture non-linear feature interactions with low inference cost (Figure 3.6).

- **Support Vector Machine (SVM)**: Kernel-based classifier using the RBF kernel, effective in non-linear decision boundaries and robust under limited training data [27].

- **Random Forest (RF)**: Ensemble of decision trees with bootstrap sampling and feature subspace partitioning, providing interpretable feature importance scores and robustness to noise [18].

All models were trained on stratified 80/20 splits with z-score normalization. Hyperparameters were tuned via Grid Search with 5-fold cross-validation, using F1-score and ROC-AUC as evaluation metrics due to class imbalance.

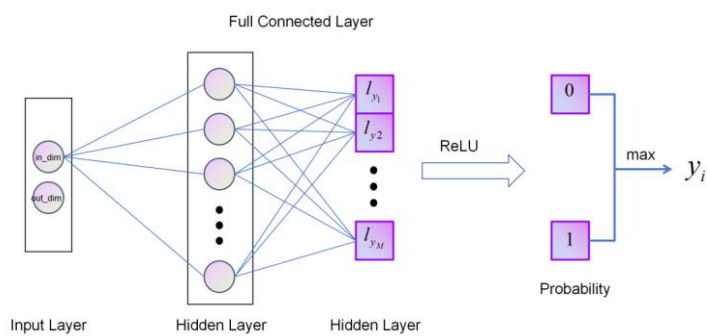

Figure 3.6: Architecture of the Multi-Layer Perceptron (MLP) used in relay classification.

### 3.3.1. MLP Results

The two-layer MLP achieved the best balance: inference latency below 1 ms, strong generalization, and stable convergence with early stopping and adaptive learning rate scheduling.

Shallow variants underperformed on minority classes, while deeper networks offered little additional benefit but higher inference cost.

### 3.3.2. SVM Results

We tested linear and RBF kernels with $C \in \{0.01, 0.1, 1, 10, 100\}$ and $\gamma \in \{0.01, 0.1, 1\}$. The optimal RBF configuration ($C = 1$, $\gamma = 0.1$) yielded an F1-score of 70.2% and ROC-AUC of 76.3%. While robust to imbalance [13], SVM suffered from longer training times due to quadratic complexity.

### 3.3.3. Random Forest Results

RF tuning varied n_estimators $\in \{50, 100, 200\}$ and max_depth $\in \{5, 10, 20\}$. The best configuration (200 trees, depth=10) achieved F1=68.4% and ROC-AUC=75.2%. Degree and contact frequency ranked highest in feature importance, consistent with prior DTN findings [28]. RF was more sensitive to small datasets and imbalance compared to MLP and SVM.

### 3.4. Flask API for Real-Time Relay Prediction

Machine learning-based routing strategies have shown strong potential to enhance DTN performance in sparse or intermittently connected scenarios. Surveys highlight that incorporating contact history and context-aware features leads to higher delivery ratios and reduced latency [29, 23, 25]. To support real-time classification, we implemented a lightweight RESTful API using Flask, decoupling model inference from the ONE simulator and enabling dynamic model updates without recompilation. The API exposes a single endpoint, /predict, which accepts features such as contact frequency, node degree, average contact duration, and latency, and returns a binary relay decision with probabilities. Pre-trained models are cached in memory, ensuring sub-millisecond inference latency, consistent with best practices in low-latency prediction serving frameworks [30].

The proposed MLPBasedSprayRouter integrates with this API to selectively forward messages. At each contact, the router extracts live features, queries the API, and forwards a copy only if the peer is classified as a high-quality relay. To reduce redun-dant calls, cached predictions are reused for repeated feature patterns, leveraging temporal locality in DTN encounters [29]. This policy enhances Spray and Wait with adaptive, data-driven relay selection, similar in spirit to MLProph [13], but with real-time inference and modular deployment.

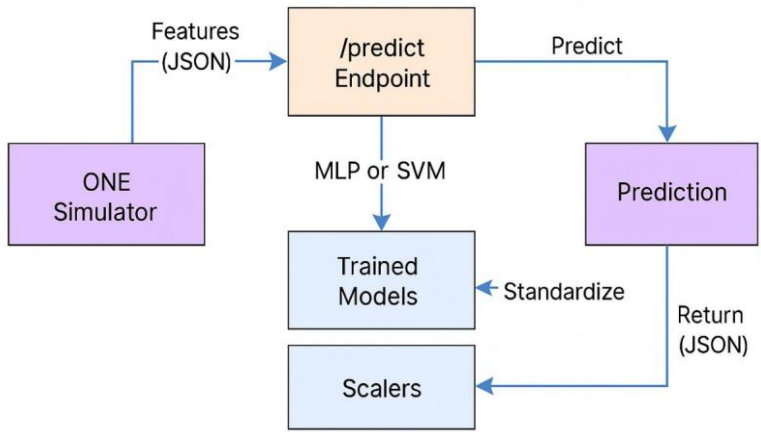

Figure 3.7: Integration of Flask API with MLPBasedSprayRouter: node features are extracted during encounters, sent to the API for classification, and used for selective forwarding decisions.

Chapter 4

# 4. Evaluation Design and Implementation

## 4.1. Scenario Design

To emulate urban emergency communication challenges, we designed a car accident scenario where conventional infrastructure is unavailable or overloaded. The simulation is based on the Helsinki downtown map [21], a widely used DTN mobility trace incorporating roads, pedestrian paths, and junctions. A stationary **accident node** periodically generates urgent messages (e.g., crash alerts, health data), which must be delivered to two fixed **hospital nodes**. Forwarding relies on heterogeneous mobile agents:

- **Pedestrian nodes**: moving at 0.5-1.5 m/s.

- **Vehicle nodes**: moving at 10-50 km/h.

Two temporal patterns were simulated: **Weekday** (structured commuter traffic) and **Holiday** (dispersed and irregular mobility) [31]. Figure 4.1 shows 12-hour contact density differences.

## 4.2. Experiment Setup

Experiments were conducted in the ONE Simulator [21]. Nine scenarios varied pedestrian and vehicle counts (PX_CY), each run for 12 hours with five random seeds.

## 4.3. Model Training and Router Integration

We compared MLP, SVM, and Random Forest classifiers on extracted features (contact frequency, degree, average duration). MLP achieved the best balance of accuracy, recall, and inference time, consistent with prior DTN ML studies [32, 28].

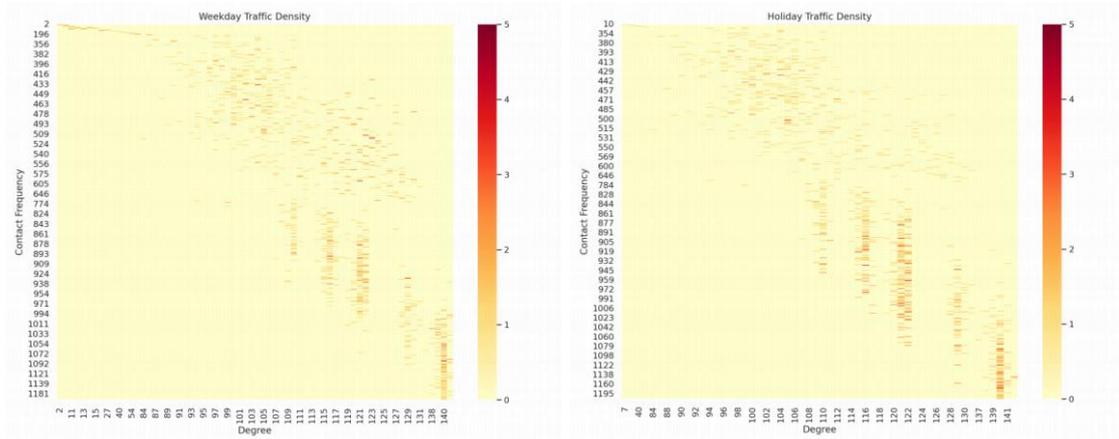

Figure 4.1: Contact density heatmaps for Weekday (left) and Holiday (right).

Table 4.1: Node configurations.

| Scenario | Pedestrian Nodes | Car Nodes |
|---|---|---|
| P50_C50 | 50 | 50 |
| P50_C60 | 50 | 60 |
| P60_C50 | 60 | 50 |
| P60_C60 | 60 | 60 |
| P70_C70 | 70 | 70 |
| P70_C80 | 70 | 80 |
| P80_C80 | 80 | 80 |
| P80_C90 | 80 | 90 |
| P90_C80 | 90 | 80 |

Table 4.2: Simulation parameters.

| | |
|---|---|
| Duration | 43,200 s (12 h) |
| Transmission range | 30 m |
| Speed | 2 Mbps |
| Buffer size | 50 MB |
| Message size | 500 KB-1 MB |
| TTL | 300 min |
| Generation interval | 25-35 s |
| Mobility model | ShortestPathMapBasedMovement |
| Map | Helsinki downtown |

Table 4.3: Model performance comparison.

| Model | Acc | Prec | Rec | F1 | AUC | Inference (s) |
|---|---|---|---|---|---|---|
| MLP | 0.763 | 0.701 | 0.782 | 0.733 | 0.775 | 0.001 |
| SVM | 0.736 | 0.709 | 0.791 | 0.731 | 0.757 | 0.023 |
| Random Forest | 0.719 | 0.703 | 0.738 | 0.704 | 0.730 | 0.014 |

The MLPBasedSprayRouter inherits from ActiveRouter, querying a Flask API for classification at each encounter. A message copy is forwarded only if the peer is predicted as a high-quality relay. This modular design enables flexible model replacement without simulator recompilation.

### 4.4. Comparison Across Temporal Scenarios

Routing performance under Weekday and Holiday conditions is summarized in Tables 4.4 and 4.5.

Table 4.4: Weekday Scenario: Protocol Results

| Metric | Spray and Wait | MLPBasedRouter | RandomRouter |
|---|---|---|---|
| Delivery Probability | 0.725 | **0.799** | 0.771 |
| Overhead Ratio | 6.653 | **9.290** | 9.385 |
| Latency Avg (s) | 3027 | **2813** | 2885 |
| Buffertime Avg (s) | 17866 | **17580** | 17787 |

Table 4.5: Holiday Scenario: Protocol Results

| Metric | Spray and Wait | MLPBasedRouter | RandomRouter |
|---|---|---|---|
| Delivery Probability | 0.717 | **0.788** | 0.765 |
| Overhead Ratio | 6.768 | **9.824** | 9.981 |
| Latency Avg (s) | 2947 | **2640** | 2695 |
| Buffertime Avg (s) | 17818 | **17620** | 17740 |

The MLPBasedRouter consistently improves delivery probability (+7-8%) and reduces latency (200-300s) in both scenarios, while buffer time is slightly reduced. Although overhead increases, the trade-off is acceptable given the delivery gains. Performance stability across weekday and holiday traffic patterns highlights the model's adaptability under dynamic urban conditions, consistent with findings in [5].

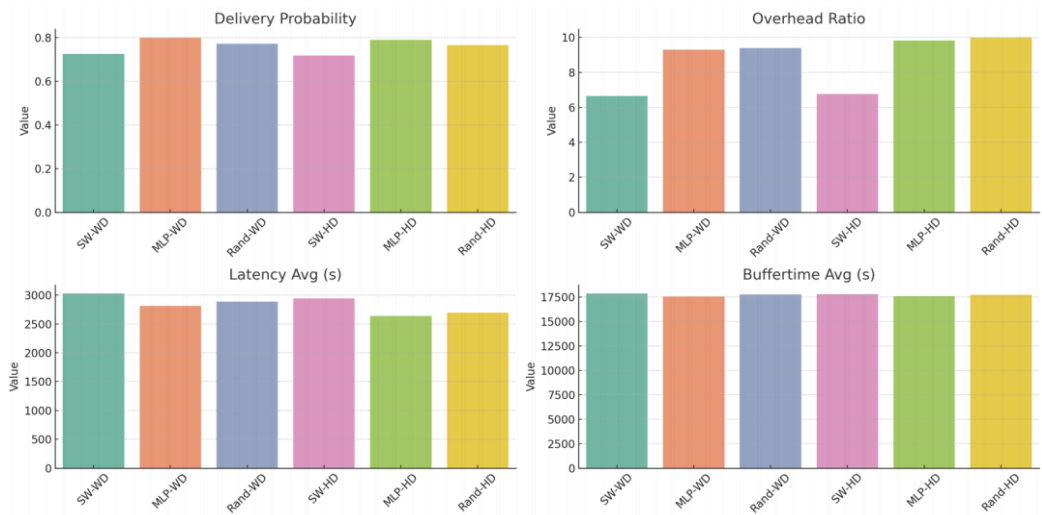

Figure 4.2: Bar chart comparison of three routing protocols (SW = Spray and Wait, MLP = MLPBasedRouter, Rand = RandomRouter) across Weekday (WD) and Holiday (HD) scenarios for four metrics: Delivery Probability, Overhead Ratio, Latency Avg, and Buffertime Avg.

Chapter 5

# 5. Discussion

Our results confirm that integrating machine learning into DTN routing provides novel and adaptive benefits under the challenging conditions of intermittent connectivity and dynamic mobility. The proposed MLPBasedSprayRouter consistently outperformed traditional Spray and Wait by improving delivery ratio and reducing latency, while maintaining competitive overhead. This aligns with prior findings that context-aware relay selection can enhance robustness in sparse and bursty environments [23, 25, 12]. The MLP classifier proved particularly effective due to its ability to capture non-linear feature interactions at low inference cost, a result consistent with other studies highlighting neural models' resilience and generalization in DTN contexts [33].

Despite these gains, several challenges remain. Performance depends heavily on the representativeness of training data and the stability of extracted features such as contact duration, degree centrality, and buffer occupancy. The use of an external inference service (Flask API) demonstrated modular integration feasibility but also raised concerns regarding scalability, caching, and inference delays in larger-scale deployments. Moreover, current models are trained offline and only partially incorporate energy- and mobility-aware dynamics, limiting adaptability in rapidly evolving scenarios.

From a practical standpoint, ML-based DTN routing shows promise for resilient communication in emergency response, vehicular networks, and rural connectivity, where conventional replication-based protocols fail to adapt. Looking ahead, future work should focus on extending feature sets, enabling online learning, and validating under real-world conditions. In particular, we plan to leverage MODiTONes [34] as a modular DTN testbed to support reproducible evaluation across diverse mobility patterns and application scenarios. This will advance the scalability, adaptability, and practicality of ML-driven DTN routing in highly dynamic environments.

Chapter 6

# 6. Conclusion and Future Work

In this paper, we presented an intelligent enhancement to the Spray and Wait protocol by integrating machine learning-based relay node selection in Delay-Tolerant Networks. Through feature extraction from simulation logs and real-time classification using lightweight models, the proposed MLPBasedSprayRouter consistently improved delivery ratio and reduced latency compared to the baseline and alternative ML approaches. Our results demonstrate the feasibility of leveraging adaptive, data-driven techniques for resilient communication in highly dynamic and resource-constrained environments such as disaster recovery and smart city deployments. Looking ahead, future work will focus on extending the feature set with energy- and mobility-aware metrics, incorporating online learning for continuous adaptation, and validating the approach in large-scale, real-world DTN testbeds. In particular, we plan to establish MODiTONes (Modular DTN Testbed for Opportunistic Networks) as a flexible and reproducible experimental platform to further evaluate ML-based routing strategies under diverse mobility patterns and application scenarios, building upon similar modular testbed efforts such as that described in [34]. These directions aim to enhance the robustness and practicality of ML-driven DTN routing for critical communication systems.